\documentclass{jfm}
\usepackage{graphicx}
\usepackage{epstopdf, epsfig}
\usepackage{amsmath}
\usepackage{amsfonts}
\usepackage{mathrsfs}
\usepackage{amssymb}
\usepackage{xcolor}
\usepackage{multicol}
\usepackage{placeins}

\makeatletter
\renewcommand*\env@matrix[1][\arraystretch]{%
  \edef\arraystretch{#1}%
  \hskip -\arraycolsep
  \let\@ifnextchar\new@ifnextchar
  \array{*\c@MaxMatrixCols c}}
\makeatother

\shorttitle{Ambiguity in mean-flow-based linear analysis}
\shortauthor{U. Karban et al.}

\title{Ambiguity in mean-flow-based linear analysis}

\author{U. Karban\aff{1}
  \corresp{\email{ugur.karban@univ-poitiers.fr}}, B. Bugeat\aff{2}, E. Martini\aff{1,4}, A. Towne\aff{3}, A.V.G. Cavalieri\aff{4}, L. Lesshafft\aff{5}, A. Agarwal\aff{2}, P. Jordan\aff{1}, T. Colonius\aff{6}}

\affiliation{
\aff{1}{D\'{e}partement Fluides, Thermique, Combustion, Institut Pprime, CNRS-University of Poitiers-ENSMA, France}
\aff{2}{Department of Engineering, University of Cambridge, Cambridge, CB2 3AP, UK}
\aff{3}{Department of Mechanical Engineering, University of Michigan, Ann Arbor, MI 48109, USA}
\aff{4}{Instituto Tecnol\'{o}gico de Aeron\'{a}utica, S\~{a}o Jos\'{e} dos Campos/SP, Brazil}
\aff{5}{Laboratoire d’Hydrodynamique, CNRS / \'{E}cole polytechnique, Institut Polytechnique de Paris, Palaiseau, France}
\aff{6}{Division of Engineering and Applied Science, California Institute of Technology, Pasadena, California 91125, USA}
}

\begin{document}

\maketitle

\begin{abstract}
Linearisation of the Navier-Stokes equations about the mean of a turbulent flow forms the foundation of popular models for energy amplification and coherent structures, including resolvent analysis.  While the Navier-Stokes equations can be equivalently written using many different sets of dependent variables, we show that the properties of the linear operator obtained via linearisation about the mean depend on the variables in which the equations are written prior to linearisation, and can be modified under nonlinear transformation of variables. For example, we show that using primitive and conservative variables leads to differences in the singular values and modes of the resolvent operator for turbulent jets, and that the differences become more severe as variable-density effects increase. This lack of uniqueness of mean-flow-based linear analysis provides new opportunities for optimizing models by specific choice of variables while also highlighting the importance of carefully accounting for the nonlinear terms that act as a forcing on the resolvent operator.  
\end{abstract}

\section{Introduction} \label{sec:intro}
Mean-flow-based linear analyses have been used since the 1970s to understand and model the dynamics of coherent structures in turbulent shear flow. The original idea behind this is that turbulence generates a mean flow that can be seen as an equivalent laminar flow on which disturbances evolve \citep{crighton_jfm_1976}. The mean flow includes some, but not all, of the effects of the non-linear flow dynamics. Approaches of this kind have evolved considerably in recent years as global stability \citep{lesshafft_jfm_2006,akervik_ejm_2008,sipp_amr_2010,mantic_prl_2014} and input-output, or resolvent analyses \citep{mckeon_jfm_2010,hwang_jfm_2010,towne_jfm_2018,schmidt_jfm_2018,cavalieri_amr_2019}, have become feasible thanks to progress in computational methods. Such analysis has been applied to a broad variety of fluid-mechanics problems (compressible, incompressible, wall-bounded and free shear flows). For the example of a turbulent jet considered later in this paper, recent reviews  have been provided by \citet{jordan_arm_2013} and \citet{cavalieri_amr_2019}. 


Many of these studies use the mean-flow-based linear operator to analyse the flow dynamics in an input-output, or resolvent, framework \citep{mckeon_jfm_2010,towne_jfm_2018}. In this case, the forcing terms are considered to drive the response through the resolvent operator. It is shown that the resolvent operator fully represents the flow dynamics if forcing is white  \citep{towne_ctr_2016,lesshafft_prf_2019}. For coloured forcing, which is generally the case for turbulent flows, flow dynamics depend also on the spectral content of the forcing \citep{zare_jfm_2017}. \cite{beneddine_jfm_2016} discussed some situations in which flow dynamics may be partially governed by the linear operator.

In other studies, eddy viscosity models have been included in the linear operator. This can enhance the extent to which the resolvent modes match the observed turbulence structure \citep{cossu_jfm_2009,hwang_jfm_2010,morra_jfm_2019,pickering_aiaa_2019}. However, physical interpretation of the remaining, unmodelled forcing terms then becomes unclear, as the use of an eddy viscosity amounts to a partial modelling of the effects of nonlinear forcing from Reynolds stress fluctuations. 


We consider mean-flow-based linear analyses from the point of view of their uniqueness. It is obvious that modifying the linear operator, either through eddy-viscosity-based modelling, or by directly changing the linearisation point, will change the characteristics of the linearised system. A less obvious ambiguity is investigated in the present paper: we aim to show that, by choosing two different, nonlinearly related, variable sets that define a given flow, one may obtain two linearised systems with different characteristics, even when the transformation between the variables is bijective. Part of this ambiguity comes from linearisation around the mean, which is a not a fixed point of the system for turbulent flows. Another part of the ambiguity is due to the non-equivalence of the means obtained for these two variable sets. A common example is the nonlinear transformation between primitive and conservative variables. We explore the effect of this choice on resolvent analysis. 

We first provide (\S\ref{sec:nstransform}) the mathematical framework that relates two linear operators obtained via nonlinear transformation of the dependent variables. We illustrate (\S\ref{sec:testcases}) the analysis by considering a number of Large-Eddy Simulation (LES) datasets for turbulent jets ranging from isothermal subsonic to heated supersonic using conservative and primitive variable sets. Based on our analysis and observations, we argue (\S\ref{sec:conc}) that the properties of the resolvent operator cannot be regarded as universal, but instead depend on the choice of variables used to define the mean, and that the overall model becomes independent of this choice only if the nonlinear forcing terms are appropriately modelled.
\section{Nonlinear transformation of Navier-Stokes equations} \label{sec:nstransform}
This study originated from an issue faced by the authors while trying to use the forcing data from an LES database constructed using a primitive-like variable set \citep{towne_phd_2016,bres_aiaa_2017,bres_jfm_2018}, with a resolvent analysis tool written in conservative variables \citep{bugeat20193d}. The problem can be illustrated by considering the effect of a nonlinear transformation of variables on the Navier-Stokes equations (though the analysis in fact applies to any nonlinear dynamical system).
\subsection{Derivation}\label{subsec:derivation}
Navier-Stokes equations (NSE) are written, 
\begin{align} \label{eq:qeq}
\partial_t \mathbf{q} = \mathcal{N}(\mathbf{q}),
\end{align}
where $\mathbf{q}\in \mathbb{R}^m$ is the vector of state variables with $m=5$ for three dimensional flows and $\mathcal{N}:\mathbb{R}^m\to\mathbb{R}^m$ is the nonlinear Navier-Stokes operator. Performing a Taylor-series expansion of $\mathbf{q}$ about a linearisation point, $\overline{\mathbf{q}}$ in state space, which is typically the mean of $\mathbf{q}$, (\ref{eq:qeq}) can be written,
\begin{align} \label{eq:qeqlin}
\partial_t \mathbf{q}^\prime -\mathcal{A}(\mathbf{q}^\prime) = \mathbf{f},
\end{align}
where $(\cdot)^\prime$ denotes fluctuations, $\mathcal{A}\triangleq D\mathcal{N}(\overline{\mathbf{q}})$ is the Fr\'{e}chet derivative of $\mathcal{N}$ at $\overline{\mathbf{q}}$ (which becomes the Jacobian $\mathbf{A}\triangleq\partial_{q}\mathcal{N}|_{\overline{\mathbf{q}}}$ for the discretised NSE) and $\mathbf{f}$ denotes all the remaining nonlinear terms, i.e., $\mathbf{f}=\mathcal{N}(\mathbf{q})-\mathcal{A}(\mathbf{q}^\prime)$. Here, an analogy with the theory of linear time-invariant (LTI) systems can be made by considering $\mathbf{A}$, as the system matrix, and $\mathbf{q}^\prime$ as the response of this system to a forcing, $\mathbf{f}$. Resolvent analysis adopts this analogy to determine a forcing function $\mathbf{f}$ that maximizes the linear gain associated with the resolvent operator, $\mathbf{R}\triangleq(i \omega \mathbf{I} - \mathbf{A})^{-1}$.

To investigate the impact of the choice of variables, we define a nonlinear bijective transformation, $\mathcal{H}:\mathbb{R}^m\to\mathbb{R}^m$, as,
\begin{align} \label{eq:trans}
\mathbf{q}_T \triangleq \mathcal{H}(\mathbf{q})
\end{align}
that maps the state variable from the original to a new set of variables, e.g., from primitive to conservative variables. The governing equation (\ref{eq:qeq}) can be represented in terms of the transformed variable $\mathbf{q}_T$ by defining another operator, $\mathcal{N}_T$ that satisfies,
\begin{align} \label{eq:veq}
\partial_t \mathbf{q}_T = \mathcal{N}_T(\mathbf{q}_T),
\end{align}
which can also be linearized around its mean, $\overline{\mathbf{q}}_T$, yielding
\begin{align} \label{eq:veqlin}
\partial_t \mathbf{q}_T^\prime -\mathcal{A}_T(\mathbf{q}_T^\prime) = \mathbf{f}_T.
\end{align}
In what follows, we express the $D\mathcal{N}_T(\mathbf{q}_T)$ in terms of $D\mathcal{N}(\mathbf{q})$. Differentiating (\ref{eq:trans}) with respect to time and inserting (\ref{eq:qeq}) and (\ref{eq:veq}) in the result gives,
\begin{align} \label{eq:vvsq}
\mathcal{N}_T(\mathbf{q}_T)=D\mathcal{H}(\mathbf{q})\mathcal{N}(\mathbf{q}),
\end{align}
where the multiplication with a Fr\'{e}chet derivative defines an inner product. To calculate the Jacobian, we need to take the derivative of the LHS of (\ref{eq:vvsq}) with respect to $\mathbf{q}_T$, which is equivalent to taking the derivative of the RHS with respect to $\mathcal{H}(\mathbf{q})$. Given that $DF(\mathcal{H}(\mathbf{q}))=D(F\circ \mathcal{H})(\mathbf{q})\left(D \mathcal{H}(\mathbf{q})\right )^{-1}$ for a smooth $F$, the Jacobian can be written as,
\begin{align} \label{eq:vvsqfinal}
D\mathcal{N}_T(\mathbf{q}_T)= D\mathcal{H}(\mathbf{q})D\mathcal{N}(\mathbf{q})\left(D \mathcal{H}(\mathbf{q})\right )^{-1} + 
 D^2\mathcal{H}(\mathbf{q}) \left(D \mathcal{H}(\mathbf{q})\right )^{-1}\mathcal{N}(\mathbf{q}).
\end{align}
The LHS of (\ref{eq:vvsqfinal}) should be calculated at $\overline{\mathbf{q}}_T$ in order to obtain $\mathcal{A}_T$. The state vectors $\overline{\mathbf{q}}$ and $\overline{\mathbf{q}}_T$ are not equivalent in the sense that they do not satisfy (\ref{eq:trans}), i.e., $\overline{\mathbf{q}}_T\neq \mathcal{H}(\overline{\mathbf{q}})$. The equivalent expansion point for the RHS of (\ref{eq:vvsqfinal}), which is not equal to $\overline{\mathbf{q}}$, is then defined as,
\begin{align} \label{eq:qtilde}
\tilde{\mathbf{q}}\triangleq\mathcal{H}^{-1}(\overline{\mathbf{q}}_T).
\end{align}
Calculation of $\overline{\mathbf{q}}_T$ requires the dynamic state data. In case only the mean flow statistics are available, a first-order prediction of $\tilde{\mathbf{q}}$ can be achieved as described in the appendix. 

Having defined $\tilde{\mathbf{q}}$, (\ref{eq:vvsqfinal}) can then be re-written as,
\begin{align} \label{eq:avsau}
{\mathcal{A}}_{T}= D\mathcal{H}(\tilde{\mathbf{q}})\tilde{\mathcal{A}}\left(D \mathcal{H}(\tilde{\mathbf{q}})\right )^{-1} + 
 D^2\mathcal{H}(\tilde{\mathbf{q}}) \left(D \mathcal{H}(\tilde{\mathbf{q}})\right )^{-1}\mathcal{N}(\tilde{\mathbf{q}}),
\end{align}
where $\tilde{\mathcal{A}}\triangleq D\mathcal{N}(\tilde{\mathbf{q}})$. Note that the second term on the RHS of (\ref{eq:vvsqfinal}) is zero if $\mathcal{H}$ is linear ($D^2\mathcal{H}(\mathbf{q})=0$ for any $\mathbf{q}$) or if linearisation is performed about a `fixed' point of the system, which, by definition, satisfies $\mathcal{N}(\overline{\mathbf{q}})=0$. In that case, the first term in (\ref{eq:avsau}) indicates a similarity transformation between ${\mathcal{A}}_T$ and $\tilde{\mathcal{A}}$, which now reduces to $\mathcal{A}$ (due to having $\overline{\mathbf{q}}_T= \mathcal{H}(\overline{\mathbf{q}})$ satisfied, by definition in case of linear $\mathcal{H}$, and through (\ref{eq:vvsq}) in case of expansion around the fixed point), and the two linear operators share the same eigenvalues. In mean-flow-based linear analysis, however, the second term on the RHS (\ref{eq:avsau}) is non-zero, which implies a difference in the eigenvalues of $\tilde{\mathcal{A}}$ and ${\mathcal{A}}_T$ even if the corresponding linearisation points are equivalent. 

We now derive an expression relating $\mathcal{A}$ and $\mathcal{A}_T$. It is more convenient to compare these two operators since one would use $\mathcal{A}$, not $\tilde{\mathcal{A}}$ when performing mean-flow linear analysis. Although (\ref{eq:avsau}) provides a means to compute $\mathcal{A}_T$, which can then be compared to $\mathcal{A}$, an exact general expression relating the two operators cannot be derived for nonlinear $\mathcal{H}$. However an approximate relation is given by replacing (\ref{eq:atilde}) into (\ref{eq:avsau}) as,

\begin{align} \label{eq:avsauapr}
{\mathcal{A}}_{T}&\approx D\mathcal{H}(\tilde{\mathbf{q}}){\mathcal{A}}\left(D\mathcal{H}(\tilde{\mathbf{q}})\right )^{-1} \nonumber \\ 
&+ D\mathcal{H}(\tilde{\mathbf{q}}) \big(\mathcal{A}D\epsilon(\overline{\mathbf{q}}) + D^2\mathcal{N}(\overline{\mathbf{q}})\epsilon(\overline{\mathbf{q}})\big)\left(D\mathcal{H}(\tilde{\mathbf{q}})\right )^{-1}
+ D^2\mathcal{H}(\tilde{\mathbf{q}}) \left(D\mathcal{H}(\tilde{\mathbf{q}})\right )^{-1}\mathcal{N}(\tilde{\mathbf{q}}).
\end{align}
The second term on the RHS of \eqref{eq:avsauapr} accounts for account the change in the linearisation point, i.e., non-equivalence of the mean flows, $\overline{\mathbf{q}}$ and $\overline{\mathbf{q}}_T$ due to taking the mean after nonlinear transformation, while the third term appears once again since the mean flow is not a fixed point of the Navier-Stokes system for turbulent flows. It is obvious that these extra terms do not cancel each other for arbitrary $\mathcal{H}$. As a result, expanded about the mean-flow, the system characteristics are modified when subjected to a nonlinear transformation. An immediate implication is that the stability characteristics of the linear operator are not unique and the linearised systems (\ref{eq:qeqlin}) and (\ref{eq:veqlin}) are equivalent only if the right-hand sides are maintained and the equal sign is respected. 
\subsection{Implications for resolvent analysis} \label{subsec:res}
Resolvent analysis involves taking the Fourier transform of (\ref{eq:qeqlin}), or similarly (\ref{eq:veqlin}), and re-organizing the result as,
\begin{align} \label{eq:res}
\hat{\mathbf{q}}=\mathbf{R}\hat{\mathbf{f}},
\end{align}
where $\hat{\mathbf{q}}$ and $\hat{\mathbf{f}}$ are Fourier transforms of $\mathbf{q}$ and $\mathbf{f}$, respectively. Ignoring the nonlinear relation between $\hat{\mathbf{f}}$ and $\hat{\mathbf{q}}$, (\ref{eq:res}) can be seen as a forcing-response relation. The optimal forcing that would maximize the energy in the response can be be found by maximising the Rayleigh quotient, $\sigma \triangleq {\langle \hat{\mathbf{q}}, \hat{\mathbf{q}} \rangle}/{\langle \hat{\mathbf{f}}, \hat{\mathbf{f}}\rangle}$, where the inner product is defined as 
\begin{align} \label{eq:inner}
\langle \mathbf{a},\mathbf{b}\rangle=\int_V\mathbf{a}^*\mathbf{M}\mathbf{b}dV,
\end{align}
with $\mathbf{M}$ defining a suitable energy norm. The Rayleigh quotient for the transformed system, $\sigma_T$, can be similarly defined using $\hat{\mathbf{q}}_T$ and $\hat{\mathbf{f}}_T$. To compare $\sigma$ and $\sigma_T$, the energy norms in the original and the transformed systems should be equivalent, which amounts to
\begin{equation} \label{eq:normequiv}
\int_V {\mathbf{q}^\prime}^*\mathbf{M}\mathbf{q}^\prime dV=\int_V {\mathbf{q}^\prime}^*_T\mathbf{M}_T\mathbf{q}^\prime_TdV.
\end{equation}
To leading order in fluctuation amplitude, $\mathbf{q}^\prime_T$ and $\mathbf{q}^\prime$ are related by the expression,
\begin{equation} \label{eq:fluc}
\mathbf{q}^\prime_T=\partial_q\mathcal{H}|_{\overline{q}}\mathbf{q}^\prime.
\end{equation}
Replacing $\mathbf{q}^\prime_T$ terms in (\ref{eq:normequiv}) using (\ref{eq:fluc}) yields a leading-order norm equivalence,
\begin{equation} \label{eq:normcorr}
\mathbf{M}=\partial_q\mathcal{H}|_{\overline{q}}^*\mathbf{M}_T\partial_q\mathcal{H}|_{\overline{q}}.
\end{equation}
Note that the norm correction is required even when the variable transformation is linear, in which case the resolvent operators are connected through a similarity transform. Equation (\ref{eq:normcorr}) can alternatively be obtained by enforcing that the Rayleigh quotients of the original and linearly transformed systems be equal. 
\section{Application to test cases} \label{sec:testcases}
The above analysis is tested using four different LES databases of ideally expanded round jets: three isothermal jets at $M_j=0.4$, 0.9 and 1.5 and a hot jet at $M_j=1.5$ are investigated to observe the effect of switching from conservative to primitive variable sets on resolvent analysis. The operating conditions are given in terms of jet exit Mach number, $M_j$, nozzle pressure ratio, $NPR\triangleq P_{0,j}/P_\infty$, nozzle temperature ratio, $NTR\triangleq T_{0,j}/T_\infty$, and Reynolds number, $Re\triangleq\rho_jU_jD/\mu_j$, where $D$ is the nozzle diameter, $\mu_j$ is the dynamic viscosity at the jet exit, and the subscript 0 is used to denote stagnation quantities. Further details about the LES databases can be found in \cite{bres_aiaa_2017,bres_jfm_2018}.  

\begin{table}
\vspace{-10px}
\caption{The list of axisymmetric jets investigated.}
\centering
\label{tab:2}
\begin{tabular}{p{2cm} c@{\hspace{.5cm}} c@{\hspace{.5cm}} c@{\hspace{.5cm}} c@{\hspace{.5cm}} c@{\hspace{.5cm}}} 
& $M_j$ & $T_j/T_\infty$ & $NPR$ & $NTR$ & $Re$ \rule{0pt}{4ex}\\
 \emph{Jet-1} & 0.4 & 1.0 & 1.117 & 1.032 & $0.45\times 10^6$ \rule{0pt}{3ex}\\ 
 \emph{Jet-2} & 0.9 & 1.0 & 1.691 & 1.162 & $1.01\times 10^6$\\ 
 \emph{Jet-3} & 1.5 & 1.0 & 3.671 & 1.450 & $1.69\times 10^6$\\ 
 \emph{Jet-4} & 1.5 & 1.79 & 3.671 & 2.596 & $0.95\times 10^6$\\ 
\end{tabular}
\end{table}

For each flow case, we performed resolvent analysis and spectral proper orthogonal decomposition (SPOD) of the axisymmetric mode using two different variable sets: $[\nu\; u_x\; u_r\; p]^T$ and $[\rho\; \rho u_x\; \rho u_r\; \rho E]^T$, which we refer to as primitive and conservative, respectively. Here, $\nu=1/\rho$, $\rho$, $u_x$, $u_r$, $p$ and $\rho E$ denote the specific volume, density, axial and radial velocities, pressure and total energy, respectively. In order to perform a singular value decomposition (SVD) of the resolvent, the Jacobian matrix of the compressible NSE, i.e., the linear operator $\mathbf{A}$ defined in \S\ref{subsec:derivation}, is computed using conservative variables by linearising the discretised equation as described by \cite{mettot2014computation}. At this point, it is possible to switch to primitive variables by first finding the linearisation point that is equivalent to the mean in primitive variables using equation \eqref{eq:qtilde}, and then applying equation \eqref{eq:avsau} to get $\mathbf{A}_T$. The SVD of the resolvent is then calculated based on Krylov methods as detailed in \cite{bugeat20193d}.
The open libraries PETSc \citep{petsc-efficient} and SLEPc \citep{Hernandez2005SSF} are used to solve the linear systems by direct LU decomposition and eigenvalue calculation by the Krylov-Schur algorithm \citep{hernandez2007krylov}.
Computations are carried out using an orthogonal mesh of $748\times 229$ points for the subsonic cases, and $636\times 229$ points for the supersonic cases, which were determined after a convergence study, in the numerical domain $x/D \in [0;30]$ and $r/D \in [0;12]$, where $x/D=0$ is the location of the nozzle exit and $r/D=0$ is the centre line.
Sponge zones are located at $x/D>20$ and $r/D>5$. 
The Chu norm \citep{chu1965energy} is used for both response and forcing over $x/D \in [0;20]$ and $r/D \in [0;5]$.

To calculate the SPOD modes, we followed the method introduced by \cite{towne_jfm_2018}. Once again, the Chu norm is used. We calculated the Fast Fourier Transform (FFT) of $\mathbf{q}$ with 128 FFT points, and overlap ratio of 0.75, yielding 310 FFT blocks for the isothermal cases, and 154 blocks for the hot jet case. 

\begin{figure}
  \centerline{\resizebox{1\textwidth}{!}{\includegraphics{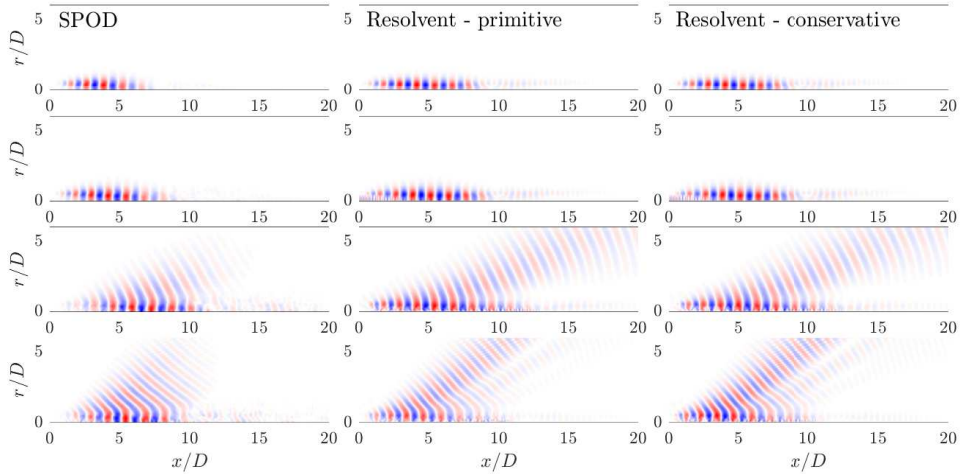}}}
  \caption{Comparison of the leading SPOD and optimal resolvent modes at $St=0.6$ for different jets (\emph{Jet-1} to \emph{Jet-4}: top to bottom). Pressure fields are shown. For the analyses using conservative variables (right), pressure is obtained through linearised transformation. }
\label{fig:optmodes}
\vspace{-15px}
\end{figure}

The optimum response modes at $St=0.6$ obtained using primitive and conservative variables, respectively, are plotted in comparison to the optimum SPOD modes in primitive variables in Figure \ref{fig:optmodes} for all jet cases. The figure shows the pressure field, which is directly computed for the primitive-variable cases, while it is reconstructed using (\ref{eq:fluc}) for the conservative-variable cases. The reconstructed SPOD modes are identical to the directly computed ones (see Table \ref{tab:5}), so are not shown here. The mode shapes are very similar between the directly computed and the reconstructed response modes. However, the global weight distribution among the variables slightly differs in \emph{Jet-4} (higher for pressure in the reconstructed mode).

\begin{figure}
  \centerline{\resizebox{1\textwidth}{!}{\includegraphics{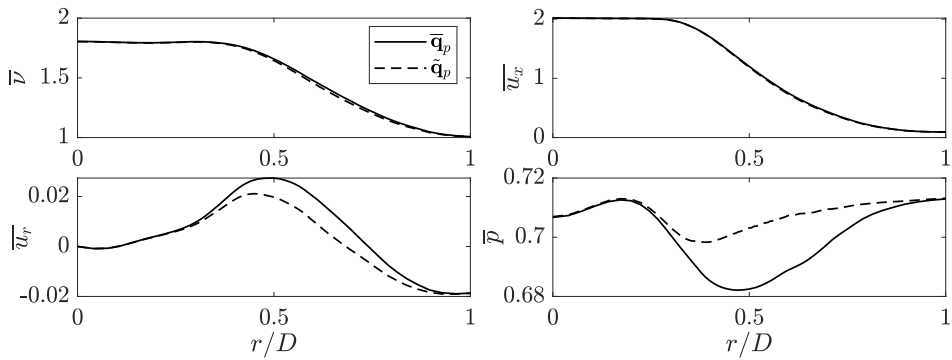}}}
  \caption{Radial mean flow profiles calculated in primitive variables (solid) and reconstructed from conservative variables (dashed) at $x/D=2$ in \emph{Jet-4}. }
\label{fig:meanx2}
\vspace{-5px}
\end{figure}

The transformation from primitive to conservative variables, which are denoted by the subscripts $p$ and $c$, respectively, is defined as 
$\mathbf{q}_c=\mathcal{H}(\mathbf{q}_p),$ with 
\begin{align} \label{eq:transfun}
\mathcal{H}(\mathbf{q})=\left[\frac{1}{\mathbf{q}_1},\;\frac{\mathbf{q}_2}{\mathbf{q}_1},\;\frac{\mathbf{q}_3}{\mathbf{q}_1},\; \frac{\mathbf{q}_4}{(\gamma-1)}+\frac{1}{2}\frac{1}{\mathbf{q}_1}\left(\mathbf{q}_2^2 + \mathbf{q}_3^2 \right)\right]^T,
\end{align}
where $\gamma=1.4$ is the heat capacity ratio. This function becomes linear for density and momentum terms when the flow is incompressible. The jet cases are listed in the order of increasing variable-density effects. Therefore we expect larger deviation in the singular values of the original and transformed resolvent operators as we move from \emph{Jet-1} to \emph{Jet-4}. To quantify the effect of density fluctuations on the mean flow, we define the percentage measure $\Delta\overline{\mathbf{q}}\triangleq\left|\mathcal{H}^{-1}(\overline{\mathbf{q}}_c) - \overline{\mathbf{q}}_p\right|/\overline{\mathbf{q}}_p\times100$. We integrate this quantity over the jet domain where turbulent kinetic energy (TKE) is greater than 1\% of its maximum value and normalize the result with the integration domain. To quantify the change in the singular values, we define, $\Delta\sigma_1 \triangleq (\sigma_{1,c}-\sigma_{1,p})/\sigma_{1,p}\times100$, and $\Delta G \triangleq (G_c-G_p)/G_p\times100$, where $G=\sigma_1/\sigma_2$ is the gain separation between the optimal and the first suboptimal singular values. The modification of the mean and the resulting changes in the singular values are tabulated in Table \ref{tab:3}. It can be seen from the table that the difference between the true mean, $\overline{\mathbf{q}}_p$ and the transformed mean, $\mathcal{H}^{-1}(\overline{\mathbf{q}}_c)$ does not exceed 0.6\% except in the radial velocity field where 7.6\% overall difference is seen in \emph{Jet-4}. The radial profiles of the mean flow calculated directly in primitive variables, $\overline{\mathbf{q}}_p$, and reconstructed from conservative variables, $\tilde{\mathbf{q}}_p$ are shown for \emph{Jet-4} in Figure \ref{fig:meanx2}, respectively. Once again, the difference between $\overline{\mathbf{q}}_p$ and $\tilde{\mathbf{q}}_p$ is mostly visible in radial velocity and pressure around the shear layer, with the largest percentage change in the former. This difference indicates a relatively strong correlation between density and radial velocity, and similarly between density and pressure, in this region. Despite the small modification of the mean, the optimal singular value and gain separation are modified by up to 40\% and 35\%, respectively, in \emph{Jet-4}.

\begin{table}
\vspace{-10px}
\caption{The change in the mean fields and the singular values calculated at $St=0.6$ due to transformation from primitive to conservative variables.}
\centering
\label{tab:3}
\begin{tabular}{p{2cm} c@{\hspace{.5cm}} c@{\hspace{.5cm}} c@{\hspace{.5cm}} c@{\hspace{.5cm}} c@{\hspace{.5cm}} c@{\hspace{.5cm}} }
& $\Delta\overline{\nu}$ [\%]& $\Delta\overline{u_x}$ [\%]& $\Delta\overline{u_r}$ [\%]& $\Delta\overline{p}$ [\%]& $\Delta\sigma_1$ [\%]& $\Delta G$ [\%] \rule{0pt}{4ex}\\
 \emph{Jet-1} & 0.00 & 0.01 & 0.31 & 0.04 & 0.05 & $-0.05$ \rule{0pt}{3ex}\\ 
 \emph{Jet-2} & 0.02 & 0.02 & 1.21 & 0.18 & $-2.81$ & $-3.25$ \\ 
 \emph{Jet-3} & 0.05 & 0.06 & 1.43 & 0.57 & $-6.02$ & $-7.45$ \\ 
 \emph{Jet-4} & 0.24 & 0.16 & 7.65 & 0.61 & $-40.58$ & $-35.87$ \\ 
\end{tabular}
\end{table} 

\begin{figure}
  \centerline{\resizebox{1\textwidth}{!}{\includegraphics{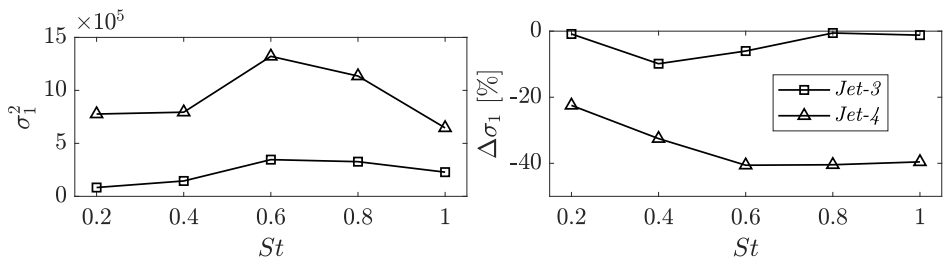}}}
  \caption{Optimal gain spectra computed using primitive variables (left), and their modification under variable transformation (right). }
\label{fig:spec}
\vspace{-10px}
\end{figure}

For the two supersonic cases, \emph{Jet-3} and \emph{Jet-4} in which noticeable changes in the singular values are observed, the analysis is repeated at different frequencies. The optimal gain spectra obtained using primitive variables and their modification under variable transformation are plotted in Figure \ref{fig:spec}. It is seen that maximum change in singular values is obtained at $St=0.4$ and $St=0.6$ for \emph{Jet-3} and \emph{Jet-4}, respectively.

We furthermore investigate the alignment of the optimum response mode of the resolvent operator with the leading SPOD mode of the response, $\mathbf{q}$. The alignment is quantified as the absolute value of the scalar product, $|\langle\mathbf{\psi}_S,\mathbf{\psi}_R\rangle|$, where the subscripts $R$ and $S$ stand for the resolvent optimal response and SPOD modes, respectively. The results at $St=0.6$ are tabulated in Table \ref{tab:4} for all jets. It is seen that the alignment is slightly better when primitive variables are used, except for \emph{Jet-4}, where the alignment is improved by $\sim$24\% when conservative variables are used, although the gain separation is lowered by 35\%. This is counter-intuitive since the alignment of SPOD and resolvent modes is often associated with high gain separation in the resolvent operator \citep{beneddine_jfm_2016}, considering that the non-linear terms are close to white noise and thus have similar projection coefficients onto the forcing mode basis. However, the presence of coherent, low-rank forcing, as recently observed for turbulent channel flow \citep{morra_arxiv_2020}, may change this picture. The overall weight of a response mode in $\hat{\mathbf{q}}$ is determined by the corresponding gain together with the projection of the actual forcing, $\hat{\mathbf{f}}$ on to the corresponding forcing mode of the resolvent operator. Therefore, the observed increase in the alignment despite of the reduction in gain separation indicates enhanced projection of forcing onto the optimal forcing mode, which is sufficient to outweigh the reduced gain separation. 

\begin{table}
\vspace{-10px}
\caption{Alignment coefficients between the leading SPOD and response modes for different jet cases at $St=0.6$. Results obtained using primitive and conservative variables, respectively, are compared.}
\centering
\label{tab:4}
\begin{tabular}{p{4.5cm} c@{\hspace{.5cm}} c@{\hspace{.5cm}} c@{\hspace{.5cm}} c@{\hspace{.5cm}} } 
& \emph{Jet-1} & \emph{Jet-2} & \emph{Jet-3} & \emph{Jet-4} \rule{0pt}{4ex}\\
 Primitive variables & 0.593 & 0.571 & 0.578 & 0.420 \rule{0pt}{3ex}\\ 
 Conservative variables & 0.587 & 0.542 & 0.555 & 0.521  \\
\end{tabular}
\rule{\textwidth}{.25pt}
\caption{Alignment coefficients between the directly computed and the reconstructed modes for different jet cases at $St=0.6$.}
\centering
\label{tab:5}
\begin{tabular}{p{4.5cm} c@{\hspace{.5cm}} c@{\hspace{.5cm}} c@{\hspace{.5cm}} c@{\hspace{.5cm}} } 
& \emph{Jet-1} & \emph{Jet-2} & \emph{Jet-3} & \emph{Jet-4} \rule{0pt}{4ex}\\
 Leading SPOD mode & 1.000 & 1.000 & 0.999 & 0.981 \rule{0pt}{3ex}\\ 
 Optimal resolvent mode & 0.996 & 0.987 & 0.983 & 0.462  \\ 
\end{tabular}
\end{table}

Finally, we investigate the change in SPOD and resolvent modes that occurs under the variable transformation. We transform the modes obtained using conservative variables to their primitive-variable counterparts using (\ref{eq:fluc}). The inner-product definition given in (\ref{eq:inner}) is now used to measure the alignment between the reconstructed and directly calculated modes. The results at $St=0.6$ are tabulated in Table \ref{tab:5}. The alignment of the leading SPOD modes calculated in conservative and primitive variables is above 98\% for all the jets. The similarity of directly computed and reconstructed optimal resolvent modes are again above 98\% for all the cases except for the heated, compressible \emph{Jet-4}, where we see a drop to 46\%. This change, consistent with the significant change observed in the singular vales, shows how the choice of variables may lead to quite different results and subsequent conclusions about the underlying flow physics. 
\section{Discussion and concluding remarks} \label{sec:conc}
We have demonstrated a non-uniqueness issue associated with mean-flow-based linear analysis. We show that the characteristics of the linear operator that is obtained by linearising a non-linear dynamical system about its mean depend on the state variables considered. In the framework of resolvent analysis, this means that two studies of a given flow, using non-linearly related dependent variables (conservative and primitive, for instance), may produce different results.


This raises questions regarding the interpretation of such analyses, now widely used to study coherent structures in turbulent shear flow. The application of such analysis for jets is often justified on the basis of their weak non-parallelism, which may result in a large gain separation between leading and sub-optimal resolvent modes, and an associated spatial separation between forcing and response modes. 
By means of resolvent analyses of turbulent jets with different operating conditions, we demonstrate how primitive- and conservative-variable-based analyses may differ significantly (up to 40\% change in gain; 35\% change in gain separation). This implies that mean-flow-based linear analyses of amplifier flows, where  flow dynamics are strongly dependent on forcing, cannot be regarded as universal, but are instead dependent on the specific form (and model) considered for the forcing terms.

A similar issue has arisen in the context of aeroacoustics. Acoustic analogies of different forms have been proposed over the years (\citet{lighthill_prs_1952,lilley1974noise,howe_jfm_1975,doak_ap_1995,goldstein_2003}; etc.) based on different linear operators considered to describe sound propagation. For each wave operator there exists a corresponding `source' term; and it is this that confounds attempts to uniquely define what is meant by a `source' of sound in turbulent flows (cf. \cite{jordan2008sja}). We see that the same situation holds for mean-flow-based stability analysis. Just as there is no unique acoustic wave propagator, there is no unique resolvent operator. However, as these approaches are all exact rearrangements of the governing equations, with residual non-linear terms  treated as external `force' or `source', all analyses will lead to the same result if the residual terms are retained. This highlights the importance of not neglecting, or over-simplifying, the external forcing term in resolvent analyis. 

While this ambiguity may be unsettling, it opens the door to optimisation of resolvent-based approaches for various applications. How to optimise the linear framework could vary depending on the problem considered. For instance, for a supersonic jet where a rank-1 model is sufficient to describe peak jet-noise \citep{sinha_jfm_2014,cavalieri_amr_2019}, finding a nonlinear transformation that maximises gain separation, whilst keeping the nonlinear forcing maximally aligned with the leading input mode, would constitute an interesting optimisation problem. For a subsonic jet, on the other hand, multiple input-output modes are necessary to correctly describe sound generation \citep{cavalieri_jfm_2014,towne_aiaa_2015,cavalieri_amr_2019}. Finding a transformation that maximises the gains and projections onto the forcing modes for the first $n$ in that case, where $n\geq2$ is to be determined as well, may help improve the modelling strategies.\\ 

This work has received funding from the Clean Sky 2 Joint Undertaking (JU) under the European Union’s Horizon 2020 research and innovation programme under grant agreement No 785303. Results reflect only the authors' view and the JU is not responsible for any use that may be made of the information it contains.
 
\vspace{-0.42cm}
	
\appendix

\section{Estimating $\tilde{\mathbf{q}}$ and $\tilde{\mathcal{A}}$ using flow statistics} \label{ap:estimmean}
Applying a Taylor series expansion to (\ref{eq:trans}) around $\overline{\mathbf{q}}$, we get
\begin{align} \label{eq:transexp}
\mathbf{q}_T = \mathcal{H}(\overline{\mathbf{q}}) + D \mathcal{H}(\overline{\mathbf{q}})\mathbf{q}^\prime + \frac{1}{2}{\mathbf{q}^\prime}^T D^2 \mathcal{H}(\overline{\mathbf{q}}){\mathbf{q}^\prime} + \mathcal{O}({\mathbf{q}^\prime}^3).
\end{align}
Neglecting third-order terms in (\ref{eq:transexp}) and taking its mean yields
\begin{align}
\overline{\mathbf{q}}_T &\approx \mathcal{H}(\overline{\mathbf{q}}) + \frac{1}{2}\overline{{\mathbf{q}^\prime}^T D^2 \mathcal{H}(\overline{\mathbf{q}}){\mathbf{q}^\prime}}.  \label{eq:transexpm}
\end{align}
Note that the first-order term in Taylor-series expansion is dropped after taking the mean, since $\overline{\mathbf{q}^\prime}=0$. Applying a linear expansion to (\ref{eq:qtilde}) again around $\overline{\mathbf{q}}$, we get
\begin{align} \label{eq:vexp}
\overline{\mathbf{q}}_T\approx\mathcal{H}(\overline{\mathbf{q}}) + D\mathcal{H}(\overline{\mathbf{q}})(\tilde{\mathbf{q}}-\overline{\mathbf{q}}).
\end{align}
(\ref{eq:transexpm}) and (\ref{eq:vexp}) both approximate $\overline{\mathbf{q}}_T$. Then, a first-order approximation of $\tilde{\mathbf{q}}$ can be obtained by rearranging these two equations as,
\begin{align} \label{eq:vapprox}
\tilde{\mathbf{q}}\approx \overline{\mathbf{q}} + \left(D \mathcal{H}(\overline{\mathbf{q}})\right)^{-1}\left( \frac{1}{2}\overline{{\mathbf{q}^\prime}^T D^2 \mathcal{H}(\overline{\mathbf{q}}){\mathbf{q}^\prime}}\right).
\end{align}
(\ref{eq:vapprox}) requires knowledge of the mean, $\overline{\mathbf{q}}$, together with the stress-like tensor, $\overline{\mathbf{q}^\prime{\mathbf{q}^\prime}^T}$. Note that, if $\mathcal{H}$ is an element-wise operator, only the mean-squares of the fluctuations, i.e., the diagonal elements of $\overline{\mathbf{q}^\prime{\mathbf{q}^\prime}^T}$ are necessary. Assuming a converged expansion in (\ref{eq:vexp}) implies that the second term in (\ref{eq:vapprox}) is small. Defining $\epsilon(\overline{\mathbf{q}})\triangleq \left(D \mathcal{H}(\overline{\mathbf{q}})\right)^{-1}\left( \frac{1}{2}\overline{{\mathbf{q}^\prime}^T D^2 \mathcal{H}(\overline{\mathbf{q}}){\mathbf{q}^\prime}}\right)$, applying $\mathcal{N}$ on both sides of (\ref{eq:vapprox}), and linearly expanding the RHS about $\overline{\mathbf{q}}$, we get, $
\mathcal{N}(\tilde{\mathbf{q}})\approx \mathcal{N}(\overline{\mathbf{q}}) + D\mathcal{N}(\overline{\mathbf{q}})\epsilon(\overline{\mathbf{q}})$, which can, finally, be differentiated with respect to $\mathbf{q}$ to obtain,
\begin{align} \label{eq:atilde}
\tilde{\mathcal{A}}\approx\mathcal{A}+\mathcal{A}D\epsilon(\overline{\mathbf{q}}) + D^2\mathcal{N}(\overline{\mathbf{q}})\epsilon(\overline{\mathbf{q}}).
\end{align} 

\vspace{-10px}
\bibliographystyle{jfm}
\bibliography{biblio}

\end{document}